\documentclass[showpacs,showkeys,preprint, preprintnumbers,amsmath,amssymb,nofootinbib, prd, aps, 11pt]{revtex4-1}
\usepackage{amsmath}
\usepackage{epsf}
\usepackage{graphicx}
\usepackage{bbm}

\newcommand{\cL}{{\cal L}}

\newcommand{\cH}{{\cal H}}
\newcommand{\cV}{{\cal V}}
\newcommand{\cE}{{\cal E}}

\newcommand{\cT}{{\cal T}}
\newcommand{\bi}{\bigskip}
\newcommand{\no}{\noindent}
\newcommand{\be}{\begin{equation}}
\newcommand{\ee}{\end{equation}}
\newcommand{\bea}{\begin{eqnarray}}
\newcommand{\eea}{\end{eqnarray}}

\newcommand{\rk}{\right)}
\newcommand{\lk}{\left(}

\newcommand{\sli}{\sum\limits}

\newcommand{\il}{\int\limits}

\newcommand{\vx}{\vec{x}}
\renewcommand{\vr}{\vec{r}}
\newcommand{\vy}{\vec{y}}

\newcommand{\vp}{\vec{p}}

\renewcommand{\vec}[1]{\mbox{\boldmath$#1$\unboldmath}}

\numberwithin{equation}{section}

\begin{document}

\title{The Cofounder of Quantum Field Theory: Pascual Jordan}

\author{W. Dittrich\\
Institut f\"ur Theoretische Physik\\
Universit\"at T\"ubingen\\
Auf der Morgenstelle 14\\
D-72076 T\"ubingen\\
Germany\\
electronic address: qed.dittrich@uni-tuebingen.de
}
%


\bi

\no

\begin{abstract}

A comparative study is undertaken that brings to light the two different methods of how to treat the many-body problem in quantum 
field theory. The two main researchers who published the first versions of how to quantize a many-body assembly were P. Jordan 
and P.A.M. Dirac. What they understood by the so-called ``second quantization'' will be the subject of the paper. We will argue that
 it is Jordan's field operator approach that until now constitutes the basis of any work in quantum field theory.

\end{abstract}


\maketitle

\section*{Introduction}

We will begin with a detailed study of the continuous string. As a simple one-dimensinal model the string with fixed end points will 
enable us to introduce the quantization of a field as function of ordinary $(x, t)$ space. Here, second quantization starts with 
an operator-field $\phi  (x, t)$. Our goal is to provide a simplified model for computing the analogue of Einstein's formula for the 
energy fluctuation of black body radiation. \cite{1} This was the starting point of Jordan's seminal contribution to quantum field theory 
(QFTh). His work was published in the last chapter of the famous ``Three- Man Paper'' (Drei-M\"anner-Arbeit) \cite{2} and for a long time 
criticized even by his coauthors. Jordan himself rated his calculation as ``almost the most important contribution I ever made to 
quantum mechanics (Q.M.).''

   It is a pity that among the founders of Q.M. Pascual P. Jordan is the least known, although he contributed more than anybody else to 
the birth of QFTh. S.S. Schweber calls him ``the unsung hero'' among the creators of Q.M.\cite{3} Just the opposite holds true for P.A.M. Dirac. 
Many experts in the field consider his paper of 1927 the beginning of quantum electrodynamics \cite{4}. While Jordan uses (like in classical field theories) 
the ordinary space-time coordinate $(\vr, t)$ as arguments of the quantum field, Dirac starts with the Schr\"odinger function with multiparticle space coordinates as argument. The second quantization is then introduced after a series of canonical transformations instead of elevating the one-particle Schr\"odinger field to an operator field that has to satisfy certain equal-time communtation relations. 
  However it was shown rather soon after Jordan's and Dirac's papers appeared that the two methods lead to identical results. Although
 the classical theories of particle and field picture are completely different as far as their mathematical and physical content is 
concerned, their quantized versions are totally identical.\cite{5} This will be demonstrated in the final chapter in modern terms. 
It remains to say that Jordan's original approach to second quantization, not Dirac's, became the standard procedure among 
researchers and in textbooks to formulate QFTh., be it non-relativistic or relativistic, which is the accepted language of 
elementary particle physics.

\section{The Classical Vibrating String}

Let us consider a vibrating string with fixed endpoints at $x = 0$ and $x = l$. Let its linear mass density be  $\sigma$  and it tension 
be $\tau$, both of which are assumed to be constant. Then the equation of motion for the vertical displacement $u  (x, t)$
 at point $x$ and time $t$ is given by
\be
\label{1.1}
\frac{1}{c^2} \frac{\partial^2 u (x,t)}{\partial t^2} = \frac{\partial^2 u (x, t)}{\partial x^2} \quad \quad
\mbox{with} \quad \quad u (0, t) = 0 = u (l, t) \, .
\ee
$c = \sqrt{\frac{\tau}{\sigma}}$ denotes the velocity in the string.

Clearly, equation (\ref{1.1}) is the wave equation in one dimension. (A derivation of equation (\ref{1.1}) 
will be given later with the aid of the action principle.)

To solve the wave equation (\ref{1.1}), let us try the Ansatz in terms of normal modes $q_n (t)$:
\be
\label{1.2}
u (x, t) = \sli^\infty_{n = 1} u_n (x) q_n (t) = \sli^\infty_{n = 1} u_n (x) c_n \cos (\omega_n t + \phi_n) \, .
\ee
To satisfy the endpoint conditions $u (0, t) = 0 = u (l, t)$                                 
we find for $u_n (x)$
\be
\label{1.3}
u_n (x) = \lk \frac{2}{l \sigma} \rk^{1/2} \sin \frac{n \pi x}{l} \, , \quad \quad n = 1, 2, \ldots \, .
\ee
$u_n (x)$ is called the n-th wave function, which satisfies the equation
\be
\label{1.4}
\lk \tau \frac{d^2}{d x^2} + \sigma \omega^2_n \rk u_n (x) = 0 \, , \quad \quad \omega_n = \frac{n \pi}{l}  c \, .
\ee
The $u_n (x)$ form a complete set of orthonormal functions:
\be
\label{1.5}
\sli^\infty_{n = 1} u_n (x) u_n (x') = \delta (x - x')
\ee
or
\begin{align}
\label{1.6}
\sli^\infty_{n = 1} \sqrt{\frac{2}{l}} \sin \frac{n \pi x}{l} \sqrt{\frac{2}{l}} \sin \frac{n \pi x'}{l}  &= \delta (x - x') \, , \nonumber\\
\il^l_0 u_n (x) \sigma u_m (x) dx &= \frac{2}{l} \il^l_0 \sin \frac{n \pi x}{l} \sin \frac{m \pi x}{l} d x \nonumber\\
&= \frac{2}{l} \frac{l}{2} \delta_{nm} = \delta_{nm} \, .
\end{align}
Incidentally, the Green's function of the string      is given by
\be
\label{1.7}
G (x, x' | \omega) =  \sli^\infty_{n = 1} \frac{u_n (x) u_n (x')}{\omega^2_n - \omega^2} = \frac{2}{\sigma l}
 \sli^\infty_{n = 1} \frac{\sin \frac{n \pi x}{l} \sin \frac{n \pi x'}{l}}{\lk \frac{c n \pi}{l} \rk^2 - \omega^2} 
\ee
with $\omega_n = \frac{n \pi}{l} \sqrt{\frac{\tau}{\sigma}} = \frac{n \pi}{l} c$ \quad or \quad
$\nu_n = n \frac{c}{2 l}$.
\bi

\no
$G (x, x' | \omega )$ given in (\ref{1.7}) solves the differential equation
\be
\label{1.8}
- \lk \tau \frac{d^2}{d x^2} + \omega^2 \sigma \rk G \lk x, x' | \omega \rk = \delta (x - x')
\ee   
with $G (0, x' | \omega) = O = G (l, x' | \omega)$.
\bi

\no
The boundary conditions are obviously satisfied; acting with $- \lk \tau \frac{d^2}{d x^2} + \omega^2 \sigma \rk$
on the Green's function solution (\ref{1.7}) leads back to the completeness relation (\ref{1.5}).
\bi

\no
Transforming back to space-time,
\be
G (x, x'; t - t') = \int \frac{d \omega}{2 \pi} e^{i \omega (t - t')}  G (x, x' | \omega) \, , \nonumber
\ee
we find
\be
\label{1.9}
G (x, x'; t - t') = \sli^\infty_{n = 1} u_n (x) \frac{\sin \omega_n  (t - t')}{\omega_n} u_n (x') \Theta (t - t') \, .
\ee
The kinetic energy of the string is obtained from the kinetic energy for an element of the string, $\frac{1}{2} (\sigma d x) \dot{u}^2$, 
and integrating over the total length $l$:
\be
\label{1.10}
T = \frac{1}{2} \il^l_0 \lk \frac{\partial u (x, t)}{\partial t} \rk^2 (\sigma d x) \, .
\ee
Using the Ansatz (\ref{1.2}) we obtain with the aid of (\ref{1.6})
\begin{align}
T &= \frac{1}{2} \sli_{n, m} \dot{q}_n \dot{q}_m \int u_n (x) \sigma u_m (x) d x \nonumber\\
&= \frac{1}{2} \sli^\infty_{n = 1} \dot{q}^2_n \, . \nonumber
\end{align}
We also need the potential energy $U = \frac{1}{2} \tau \il^l_0 \lk \frac{\partial u (x, t)}{\partial x} \rk^2 d x$
or, after an integration by parts and fixed endpoints, $u (0, t) = 0 = u (l, t)$
\be
\label{1.11}
U = - \frac{1}{2} \tau \il^l_0 u (x, t) \frac{\partial^2 u (x, t)}{\partial x^2} d x \, .
\ee
Again using the Ansatz (\ref{1.2}) and equation (\ref{1.4}), i.e.,
\begin{align*}
\frac{d^2}{d x^2} u_n (x) &= - \frac{\sigma}{\tau} \omega^2_n u_n (x) \, , \\
U &= - \frac{1}{2} \tau \sli_{n, m} q_n (t) q_m (t) (- \omega^2_n) \frac{1}{\tau} \il^l_0 u_n (x) \sigma u_m (x) d x \\
&= \sli^\infty_{n = 1} \omega^2_n q^2_n (t) \, .
\end{align*}
Therefore the Lagrangian is given by
\be
\label{1.12}
L = T - U = \frac{1}{2} \sli^\infty_{n = 1} \lk \dot{q}^2_n - \omega^2_n q^2_n \rk \, , \quad \quad \omega_n = \frac{n \pi}{l} c \, .
\ee
This result is quite remarkable. The normal coordinates reduce the Lagrangian to a discrete set of uncoupled harmonic oscillators. 
The equations of motion for the normal coordinates are
\be
\label{1.13}
\ddot{q}_n + \omega^2_n q_n = 0 \, , \quad \quad n = 1, 2, \dots, \infty \, .
\ee
The total energy is given by $E = T + U$.
If we substitute $q_n (t) = c_n \cos \lk \omega_n t + \phi_n \rk$, then we obtain
\begin{align*}
T &= \frac{1}{2} \sli^\infty_{n = 1} \dot{q}_n (t)^2 = \frac{1}{2} \sli^\infty_{n = 1} \omega^2_n \lk - c_n \sin \lk \omega_n t + \phi_n \rk \rk^2 \nonumber\\
U &= \frac{1}{2} \sli^\infty_{n = 1} \omega^2_n q_n (t)^2 = \frac{1}{2} \sli^\infty_{n = 1} \omega^2_n \lk c_n \cos \lk \omega_n t + \phi_n \rk \rk^2 \, .
\end{align*}
Hence the total energy E of the string is a constant of motion:
\be
\label{1.14}
E = T + U = \frac{1}{2} \sli^\infty_{n = 1} \omega^2_n c^2_n \, .
\ee
Again, it is quite remarkable that the total energy of the vibrating string is given by a sum of contributions from each of the normal modes.

To formulate Hamilton's principle for the string we need a Lagrangian, so that we can perform the variation:
\begin{align*}
W_{12} & = \il^{t_1}_{t_2} d t L \, , \\
\delta W_{12} &= G [t_1] - G [t_2] \, ,
\end{align*}
where
\be
\label{1.15}
L (t) = \il^l_0 d x {\cal L} 
\lk \varphi (x, t), \frac{\partial \varphi (x, t)}{\partial t} 
\, , \, \frac{\partial \varphi (x, t)}{\partial x} \, ; \, x, t \rk \, . 
\ee
(Notice that we replaced $u (x, t)$ by $\varphi (x, t)$.)

The Lagrangian density $\cL$ is itself a function of the ``field'' $\varphi (x, t)$.
 The role of the generalized velocities is taken over by $\frac{\partial \varphi (x, t)}{\partial t}$.The dependence
on $\frac{\partial \varphi (x, t)}{\partial x}$ is new; it represents an ``interaction''
 between the infinitesimally separated elements of the system. Note that $\cL$ has the dimension of energy per unit length.

In general the generalized coordinate $\varphi (x, t)$ is varied according to
\be
\varphi (x, t) \to \varphi (x, t) + \delta \varphi (x, t) \nonumber
\ee
with the spatial boundary condition $\delta \varphi (x = 0, t) = 0 = \delta \varphi (x = l, t)$.
We will, however, also allow for the time variation at the endpoints:
\be
\delta \varphi (x, t_1) \neq 0 \neq \delta \varphi (x, t_2) \, , \quad \quad  \mbox{for all} \quad x. \nonumber
\ee
Hamilton's principle then takes the form
\be
\delta W_{12} = \il^{t_1}_{t_2} d t \lk L (t) \frac{d}{dt} (\delta t) + \delta L \rk \, , \nonumber
\ee
where
\be
L (t) = \il^l_0 d x \cL \nonumber
\ee
with 
\begin{align}
\label{1.16}
\cL &= \frac{\sigma}{2} \lk \frac{\partial \varphi}{\partial t} \rk^2 - \frac{\tau}{2} \lk \frac{\partial \varphi}{\partial x} \rk^2 \, , 
\quad \quad \sigma, \tau = const. \nonumber\\
 &= \cT- \cV \, .
\end{align}
Using
\be
\delta \cL = \frac{\partial \cL}{\partial t} \delta t + \frac{\partial \cL}{\partial x} \delta x + \frac{\partial \cL}{\partial \varphi} \delta \varphi + 
\frac{\partial \cL}{\partial \lk \frac{\partial \varphi}{\partial t} \rk} \delta \lk \frac{\partial \varphi}{\partial t} 
\rk + \frac{\partial \cL}{\partial \lk \frac{\partial \varphi}{\partial x} \rk } \delta \lk \frac{\partial \varphi}{\partial x} \rk \nonumber
\ee
with
\be
\frac{\partial \cL}{\partial \varphi} = 0 \, , \quad \quad \frac{\partial \cL}{\partial \lk \frac{\partial \varphi}{\partial t} \rk} 
= \sigma \lk \frac{\partial \varphi}{\partial t} 
\rk \, , \quad \quad \frac{\partial \cL}{\partial \lk \frac{\partial \varphi}{\partial x} \rk } = - \tau \lk \frac{\partial \varphi}{\partial x} \rk \nonumber
\ee
we obtain
\be
\delta \cL = \sigma \frac{\partial \varphi}{\partial t} \frac{\partial}{\partial t} (\delta \varphi) - \sigma \lk \frac{\partial \varphi}{\partial t} \rk^2 
\frac{\partial}{\partial t} (\delta t) - \tau \frac{\partial \varphi}{\partial x} \frac{\partial }{\partial x} (\delta \varphi) \, .
\nonumber
\ee
After a few more steps we finally end up with the Lagrangian formulation:
\begin{align}
\label{1.17}
\delta W_{12} &= \il^{t_1}_{t_2} d t \il^l_0 d x \left\{ \frac{\partial}{\partial t} \left[ \sigma \lk \frac{\partial \varphi}{\partial t} \rk \delta \varphi
 - \lk \frac{\sigma}{2} \lk \frac{\partial \varphi}{\partial t} \rk^2 + \frac{\tau}{2} \lk \frac{\partial \varphi}{\partial x} \rk^2 \rk d t \right]
\right. \nonumber\\
& \left. - \lk \sigma \frac{\partial^2 \varphi}{\partial t^2} - \tau \frac{\partial^2 \varphi}{\partial x^2} \rk \delta \varphi + \delta t 
\frac{\partial}{\partial t}
\left[ \frac{\sigma}{2} \lk \frac{\partial \varphi}{\partial t} \rk^2 + \frac{\tau}{2} \lk \frac{\partial \varphi}{\partial x} \rk^2 \right] \right\}
\end{align}
From here we can identify: $(\delta W_{12} = 0)$
\begin{align}
\label{1.18}
\delta \varphi: & \hspace{1cm} \sigma \frac{\partial \varphi}{\partial t^2} - \tau \frac{\partial^2 \varphi}{\partial x^2} = 0 \quad \quad
& \mbox{equation of motion, i.e., one-}\nonumber\\
& & \mbox{dimensional string equation} \, . \nonumber\\
\delta t: & \hspace{1cm}  \frac{\partial}{\partial t} \left[ \frac{\sigma}{2} \lk \frac{\partial \varphi}{\partial t} \rk^2 + \frac{\tau}{2}  \lk 
\frac{\partial \varphi}{\partial x} \rk^2 \right] = 0 \quad \quad 
& \mbox{energy conservation}
\end{align}
or
\be
\label{1.19}
\frac{\partial}{\partial t} (\cT + \cV =  \cE) =  0 \, .
\ee
Surface term                                                                                    
\be
\label{1.20}
G = \il^l_0 d x \left[ \sigma \lk \frac{\partial\varphi}{\partial t} \rk \delta \varphi - \cE \delta t \right]
\ee
with
\be
\label{1.21}
\Pi = \sigma \frac{\partial \varphi}{\partial t} \, , \quad \quad \mbox{momentum density} 
\ee
and 
\begin{align}
\label{1.22} 
\cE &= \frac{\sigma}{2} \lk \frac{\partial \varphi}{\partial t} \rk^2 + \frac{\tau}{2} \lk \frac{\partial \varphi}{\partial x} \rk^2 \, ,
\mbox{total energy density} \nonumber\\
E &= \il^l_0 d x \cE \, .
\end{align}
The momentum density conjugate to  $\varphi (x, t)$ is defined as
\be
\label{1.23}
\Pi (x, t) = \frac{\partial \cL}{\partial \lk \frac{\partial \varphi}{\partial t} \rk} (= \sigma \frac{\partial \varphi}{\partial t} \quad 
\mbox{for the string}) \, .
\ee
The Hamiltonian density is in general given by
\be
\cH \lk \varphi, \Pi, \frac{\partial \varphi}{\partial x}; x, t \rk = \Pi \frac{\partial \varphi}{\partial t} - \cL \, . \nonumber
\ee
For our one-dimensional string this yields with (\ref{1.16}) and (\ref{1.23}):
\begin{align*}
\cH &= \sigma \frac{\partial \varphi}{\partial t} \frac{\partial \varphi}{\partial t} - \lk \frac{\sigma}{2} \lk \frac{\partial \varphi}{\partial t} \rk^2
- \frac{\tau}{2} \lk 
\frac{\partial \varphi}{\partial x} \rk^2 \rk = \frac{\sigma}{2} \lk \frac{\partial \varphi}{\partial t} \rk^2 + \frac{\tau}{2} \lk 
\frac{\partial \varphi}{\partial x} \rk^2 \nonumber\\
&= \frac{1}{2 \sigma} \lk \Pi (x, t) \rk^2 + \frac{\tau}{2} \lk \frac{\partial \varphi}{\partial x} \rk^2 = \cT + \cV \,  \nonumber
\end{align*}
and therefore										    
\be
\label{1.24}
H (t) = \il^l_0 d x \cH (x, t) \, .
\ee
In the Hamiltonian formulation we obtain:
\begin{align}
\label{1.25}
\delta W_{12} &= \il^1_2 d t \il^l_0 d x \left\{ \frac{\partial}{\partial t} \left[ \Pi \delta \varphi - \cH \delta t \right] \right. \nonumber\\
& + \delta \varphi \lk - \frac{\partial \Pi}{\partial t} - \frac{\partial \cH}{\partial \varphi} + \frac{\partial}{\partial x} 
\frac{\partial \cH}{\partial \! \lk  \frac{\partial \varphi}{\partial x} \rk} \rk \nonumber\\
& \left. + \delta \Pi \lk \frac{\partial \varphi}{\partial t} - \frac{\partial \cH}{\partial \Pi}  \rk + \delta t \frac{\partial}{\partial t} \cH \right\} 
\end{align}
so that $\delta W_{12} = 0$ implies
\begin{align}
\label{1.26}
\delta \varphi: \frac{\partial \Pi}{\partial t} &= -  \frac{\partial \cH}{\partial \varphi} + \frac{\partial}{\partial x}
\frac{\partial \cH}{\partial \! \lk \frac{\partial \varphi}{\partial x} \rk} \, , \nonumber\\
\delta \Pi: \frac{\partial \varphi}{\partial t} &= \frac{\partial \cH}{\partial \Pi} \, , \nonumber\\
\delta t: \frac{\partial \cH}{\partial t} &= 0 \, .
\end{align}
Surface term:
\begin{eqnarray}
\label{1.27}
G &= & \il^l_0 d x \lk \Pi (x, t) \delta \varphi (x, t) - \cH \delta t \rk \, , \\
\label{1.28}
\cH &=& \frac{1}{2 \sigma} \lk \Pi (x, t) \rk^2 + \frac{\tau}{2} \lk \frac{\partial \varphi}{\partial x} \rk^2 \, .
\end{eqnarray}
\bi

\no
\section{Pascual Jordan's field quantization of the string - the origin of quantum field theory}

After the introduction to classical Hamiltonian mechanics for fields, we can now begin to quantize our so far classical string field 
according to canonical quantization. For this reason Jordan replaced the classical fields $\varphi (x, t), \Pi (x, t)$
  etc. by quantized operator fields which have to satisfy certain equal-time commutation relations. To derive these relations I 
will state without proof the following fundamental operator statement of the quantum action principle:
Suppose we have an operator  $O = O (\varphi (t))$
 and we consider a variation corresponding to a change of the parametric variable $t$; then the infinitesimal change of the operator $O$ is given by
\be
\label{2.1}
\delta O = - \frac{i}{\hbar} [O, G] \, .
\ee
Let at a given time $(\delta t = 0)$ the variation $\delta \varphi (x, t) \neq 0$; then the operator  version of (\ref{1.27}) is given by
\be
G (t) = \il^l_0 d x' \Pi (x', t) \delta \varphi (x', t) \, . \nonumber
\ee
Now, using (\ref{2.1}) with $O = \varphi (x, t)$, we have:
\be
\delta \varphi (x, t) = - \frac{i}{\hbar} \left[ \varphi (x, t), \il^l_0 d x' \Pi (x', t) \delta \varphi (x', t) \right] \nonumber
\ee
or 												     
\be
\label{2.2}
\left[ \varphi (x, t), \Pi (x',  t) \right] = i \hbar \delta (x - x') \, .
\ee
Here is a list of Jordan's equal-time commutation relations in the Heisenberg picture:
\begin{align}
\label{2.3}
[\varphi (x, t), \varphi (x', t) &= 0 \, , & [q_i, q_j] &= 0 \, , \nonumber\\
[\Pi (x, t), \Pi (x', t)] &= 0 \, , & [p_i, p_j] &= 0 \, , \nonumber\\
[\varphi (x, t), \Pi (x', t)] &= i \hbar \delta (x- x') \, , & [q_i, p_j] &= i \hbar \delta_{ij} \, .
\end{align}
The discrete analogies are on the right. For the string field we found
$\Pi (x, t) = \sigma \frac{\partial \varphi}{\partial t}$.

In the following we want to employ our results obtained so far to compute the mean-square energy fluctuation in a narrow frequency interval 
$(\omega, \omega + \Delta \omega)$ in a small segment of the string, i.e., in the region
 $0 \leq x \leq a, a \ll l$. We follow mainly the paper by the author of ref. \cite{6}.
Our former $\varphi (x, t)$ is from now on considered a quantized field. Again we expand in slightly modified modern notation:
\be
\label{2.4}
\varphi (x, t) = \sli^\infty_{n = 1} \lk \frac{\hbar}{\omega_n} \rk^{1/2} q_n (t) u_n (x) \, , \mbox{where, as before} \, ,
u_n (x) = \lk \frac{2}{\sigma l} \rk^{1/2} \sin k_n x
\ee
with
\be
\label{2.5}
q_n (t) = \frac{1}{\sqrt{2}} \lk a_n e^{- i \omega_n t} + a^\dagger_n e^{i \omega_n t} \rk \, , \quad \omega_n = c k_n = n 
\lk \frac{c \pi}{l} \rk \, .
\ee
At this point it would be useful to say a few words about the Hermitean particle operator $N =  a^\dagger a$
  and the creation and annihilation operators $a^\dagger$   and $a$ 
 of the excitations of the normal modes of the string. It will also help us to better understand Dirac's definition of of 
$b^\dagger_r$  and $b_r$  in terms of the amplitude $(\sqrt{N_r})$ and phase $(\Theta_r)$. 
The great idea of Dirac was to make the Schr\"odinger wave function a multiparticle function of particle numbers 
$N'_r, \psi (N'_1, N'_2, \ldots, N'_r, \ldots, t)$  and show how this function (not state)
 is changed when acted upon by the creation and destruction operator $b_r$  and $b^\dagger_r$.
Let us summarize some known facts from oscillator physics:
\begin{align}
\label{2.6}
N & = a^\dagger a = N^\dagger \, , \quad [a, a^\dagger] = 1 \nonumber\\
[a, N] &= a a^\dagger a - a^\dagger a a = \lk a a^\dagger - a^\dagger a \rk a = [a, a^\dagger] a = a \, .
\end{align}
So we obtain
\begin{align}
\label{2.7}
a N - N a = a \, , \nonumber\\
\mbox{or} \quad a N (N + 1) a \, , \nonumber\\
\mbox{or} \quad a (N - 1) = N a \, .
\end{align}
Likewise for the adjoint $a^\dagger$:
\begin{align}
\label{2.8}
a^\dagger N - N a^\dagger &= - a^\dagger \, , \nonumber\\
\mbox{or} \quad N a^\dagger &= a^\dagger (N + 1) \, , \nonumber\\
\mbox{or} \quad (N - 1) a^\dagger &= a^\dagger N \, .
\end{align}
Starting with  $a N = (N + 1) a$
we obtain immediately $a N^2 = (N + 1) a N = (N + 1)^2 a, \ldots$ etc.

Hence, for every function that can be expanded in a power series, we have
\be
\label{2.9}
a f (N) = f (N + 1) a 
\ee
an example of which is 
\be
\label{2.10}
a^{- \lambda N} = e^{- \lambda (N + 1)} a \, ,
\ee
 where $\lambda$   is a parameter. The proof for this important result is simple:
\begin{align*}
e^{\lambda N} a e^{- \lambda N} &= a + \lambda [N, a] + \frac{\lambda^2}{2!} [N, [N, a]] + \cdots \\
&= \lk 1 - \frac{\lambda}{1!} + \frac{\lambda^2}{2!} - \frac{\lambda^3}{3!} + \cdots \rk a = e^{- \lambda} a \, .
\end{align*}
Therefore
\be
a e^{- \lambda N} = e^{- \lambda N} e^{- \lambda} a )= e^{- \lambda (N + 1)} a \, .\nonumber
\ee
Dirac's operators $b_r$   are defined in terms of the aforesaid amplitude and angle variables:
\be
\label{2.11}
b_r = e^{- \frac{i}{\hbar} \Theta_r} \sqrt{N_r} \, , \quad [\Theta_s, N_r] = i \hbar \delta_{rs} \, ,
\ee
whence one finds, using the representation $\Theta_r = i \hbar \frac{\partial}{\partial N_r}$, and consequently 
$e^{\frac{i}{\hbar} \Theta_r} = e^{- \frac{\partial}{\partial N_r}}$ 
will decrease $N_r$  by unity,
whereas $e^{- \frac{i}{\hbar} \Theta_r} = e^{\frac{\partial}{\partial N_r}}$   will increase $N_r$  by unity.	             

Thus, setting $a \equiv e^{\frac{\partial}{\partial N_r}}$ in (\ref{2.9}), we find:
\be
\label{2.12}
e^{- \frac{i}{\hbar} \Theta_r} f (N_r) = f (N_r + 1) e^{- \frac{i}{\hbar} \Theta_r} \quad \mbox{or} \quad e^{\frac{\partial}{\partial N_r}} f (N_r)
= f (N_r + 1) e^{\frac{\partial}{\partial N_r}} \, .
\ee
This brings us to the following set of useful formulae:
\begin{align}
\label{2.13}
b_r &= e^{- \frac{i}{\hbar} \Theta_r} \sqrt{N_r} = \sqrt{N_r + 1} e^{- \frac{i}{\hbar} \Theta_r} = e^{\frac{\partial}{\partial N_r}}
 \sqrt{N_r} = \sqrt{N_r + 1} e^{\frac{\partial}{\partial N_r}} \, ,  \nonumber\\
b^\dagger_r & = \sqrt{N_r} e^{\frac{i}{\hbar} \Theta_r} = e^{\frac{i}{\hbar} \Theta_r} \sqrt{N_r + 1} = \sqrt{N_r} 
e^{- \frac{\partial}{\partial N_r}} = e^{- \frac{\partial}{\partial N_r}}
\sqrt{N_r + 1} \, , \nonumber\\
N_r &= b^\dagger_r b_r \, .
\end{align}
with $N_r$ equal to zero or a positive integer for the number of particles in the state $r$.

Let's pick up the story at (\ref{2.5}) and add to $q_n (t)$ the dimensionless conjugate momentum operator
\be
\label{2.14}
p_n 
 (t) = \frac{1}{\sqrt{2} i}  \lk a_n e^{- i \omega_n t} - a^\dagger_n e^{i \omega_n t} \rk \, .
\ee
Then the Born-Jordan commutation relations in dimensionless form are given by
\be
\label{2.15}
[a_n, a^\dagger_m] = \delta_{nm} \, , \quad \quad [q_n (t), p_m (t)] = i \delta_{nm} \, .
\ee
The transition from the quantum matter-field to the total energy operator of the string is given by
\begin{align}
\label{2.16}
H &= \frac{1}{2} \il^l_0 d x \left[ \sigma \lk \frac{\partial \varphi}{\partial t} \rk^2 + \tau \lk \frac{\partial \varphi}{\partial x}
\rk^2 \right] = \frac{1}{2} \sli^\infty_{n = 1} \hbar \omega_n \lk a^\dagger_n a_n + a_n a^\dagger_n \rk \nonumber\\
&= \sli^\infty_{n = 1} \hbar \omega_n \lk a^\dagger_n a_n + \frac{1}{2} \rk \, .
\end{align}
With the following replacements of the operators by $c$-numbers
\be
a_n \to \lk \frac{\omega_n}{2 \hbar} \rk ^{1/2} c_n e^{- i \phi_n} \, , \quad \quad a^\dagger_n \to \lk \frac{\omega_n}{2 \hbar} \rk^{1/2} c_n e^{i \phi_n} \nonumber
\ee
we recover the classical expression (\ref{1.14}) 
for the total energy of the string. The real classical dynamical variables take the form
\begin{align}
\lk \frac{\hbar}{\omega_n} \rk^{1/2} q_n (t) \to q^{cl}_n (t) & = c_n \cos (\omega_n t + \phi_n) \, , \nonumber\\
\lk \frac{\hbar}{\omega_n} \rk^{1/2} p_n (t) \to p^{c 1}_n (t) &= - \omega_n c_n \sin (\omega_n t + \phi_n) = \dot{q}^{cl}_n (t) \, . \nonumber
\end{align}
Now we are in a position to present a short introduction  to the calculation of the mean-square energy fluctuation for the quantized string field.

Let us first calculate the energy $E_{a, \omega}$         of a piece of length a of the string in the narrow frequency range $\omega, \omega + \Delta \omega$:
\be
E  (t) = \frac{1}{2} \il^a_0 d x \left[\sigma \lk \frac{\partial \varphi}{\partial t} \rk^2 + \tau \lk \frac{\partial \varphi}{\partial x} \rk^2 
\right] \, .
\nonumber
\ee
One obtains with $H$ from (\ref{2.16}):
\begin{align}
\label{2.17}
E_{a, \omega} (t) &= \frac{a}{l} H + \frac{1}{2 l} \sli_{n \neq m} \hbar \sqrt{\omega_n \omega_m} \left[ q_n q_m K'_{nm} + p_n p_m K_{nm} \right] \nonumber\\
& + \sli_n \hbar \omega_n \lk q^2_n - p^2_n \rk \sin (2 k_n a) \frac{1}{2 k_n} \, , \quad \quad \omega_n = c k_n   = \lk \frac{c \pi}{l} \rk
n \, , \quad \lambda_n = \frac{2 l}{n} \, .\\
\label{2.18}
K'_{nm} &= \frac{\sin (k_n - k_m) a}{k_n - k_m} + \frac{\sin (k_n + k_m) a}{k_n + k_m} \, , \nonumber\\
K_{nm} &= \frac{\sin (k_n - k_m) a}{k_n - k_m} - \frac{\sin (k_n + k_m) a}{k_n + k_m} \, .
\end{align}
Since we will be interested in the fluctuations in a narrow spectral range with $\lambda_n = \frac{2 l}{n}$
 much smaller than both $l$ and $a$, the final term in (\ref{2.17}) 
can be neglected due to fast oscillations, so the the deviation of the energy from the diagonal part, $\frac{a}{l} H$ in (\ref{2.17}), is given by
\be
\label{2.19}
\Delta E_{a, \omega} = \frac{1}{2 l} \sli_{n \neq m} \hbar \sqrt{\omega_n \omega_m} \left[ q_n q_m KÄ'_{nm} + p_n p_m K_{nm} \right] \, .
\ee
From here on the reader can follow the explicit calculations outlined in chapter 5.2 of ref. \cite{6}. 
There is also Jordan's original calculation in the Drei-M\"anner-Arbeit of Born, Heisenberg and Jordan \cite{2}. 
Furthermore, it is rewarding to read up on the string field history in Heisenberg's Chicago lecture of 1929 \cite{7} 
and Jordan's conference contribution in Charkow in May of 1929 \cite{8}.

The result of Jordan's work is
\be
\label{2.20}
\overline{\Delta E^2_{a, \nu}} = \frac{c}{2 a \Delta \nu} \overline{E}^2_{a, \nu} + h \nu \overline{E}_{a, \nu} \, .
\ee
in complete analogy with Einstein's mean square energy fluctuation formula for black body radiation:
\be
\label{2.21}
\overline{\Delta E^2_{V, \nu}} = \frac{c^3}{8 \pi \nu^2 V \Delta \nu} \overline{E}^2_{V, \nu} + h \nu \overline{E}_{V, \nu} \, .
\ee
In either case, the origin of the particle-like fluctuation, second part in (\ref{2.20}), 
is the presence of the zero-point energy term due to the non-commutativity of the quantized amplitudes $q_n (t)$ and $p_m (t)$. 
The appearance of this particle-like fluctuation is a kinematic effect which arises from the quantized nature of the string (photon) field. 
Jordan himself considered his fluctuation calculation as ``almost the most important contribution I made to quantum mechanics.'' In fact, 
it was the beginning of non-relativistic quantum field theory, if only for free fields. However, very shortly thereafter Dirac started his 
first attempt to formulate quantum electrodynamics. Nevertheless, Jordan introduced for the first time the quantum field operator (second quantization)
 as equivalent description to many particle wave functions. This is until the present day the accepted language for treating relativistic particle 
processes - with all its inherent problems that stem from products of relativistic local quantum field 
operators.

\section{Dirac's paper on ``The Quantum Theory of the Emission and Absorption of Radiation''}

Jordan's contribution in the final section of the Drei-M\"anner-Arbeit from November 15, 1925, \cite{2}
 and Dirac's paper from February 2, 1927, \cite{4} contain the foundation of quantum field theory, in particular of quantum electrodynamics. However, 
it is not correct to say that Dirac also invented the so-called second quantization procedure as we understand it today. 
Second quantization in its modern meaning as treating effectively many particle problems with the aid of quantum field operators is Jordan's 
contribution. Dirac baptized his procedure in this way because he introduced a pair of canonical variables, $b_r$ and $b^\dagger_r$, 
which satisfy commutation rules and then act on a many-particle Schr\"odinger wave function, while Jordan elevated the one-particle Schr\"odinger
 wave function to charged matter-field operators $\psi (\vr,  t), \psi^\dagger (\vr, t)$ of ordinary space-time.
As ``second quantized'' field operators the  $\psi (\vr, t)$ and $\psi^\dagger (\vr, t)$
 satisfy certain equal-time commutation relations. We have already seen in chapter 2 how these follow from the quantum action principle. 
It is only necessary to change the generator G(t) of the string theory to the appropriate form for the Schr\"odinger theory. 
Without further proof, the momentum conjugate to  $\psi (\vr, t)$ is given by $\Pi (\vr, t) = i \hbar \psi^\dagger (\vr, t)$ so that
\be
\label{3.1}
G (t) = \int d^3 \vr' i \hbar \psi^\dagger (\vr', t) \delta \psi (\vr' t) \, ,
\ee
which yields
\be
\label{3.2}
\delta \psi (\vr, t) = - \frac{i}{\hbar} [\psi (\vr, t), \int d^3 \vr' i \hbar \psi^\dagger (\vr', t) \delta \psi (\vr', t)] \, .
\ee
From here we obtain
\be
[\psi (\vr, t), \psi^\dagger (\vr', t)] = \delta (\vr - \vr') \, . \nonumber
\ee
We also add the relations
\begin{align}
\label{3.3}
[\psi (\vr, t), \psi (\vr', t)] &= 0 \, , \nonumber\\
[\psi^\dagger (\vr, t), \psi^\dagger (\vr', t)] &= 0 \, .
\end{align}
Expanding $\psi$ and $\psi^\dagger$ in terms of a system of orthogonal c-number functions $\varphi_r (\vr)$, 
\be
\label{3.4}
\psi (\vr, t) = \sli_r b_r (t) \varphi_r (\vr) \, , \, \psi^\dagger (\vr, t) = \sli_\nu b^\dagger_r (t) 
\varphi^*_r (\vr) \, , \,
 \int d^3 \vr \varphi_r (\vr) \varphi^*_s (\vr) = \delta_{r, s} \, ,
\ee
the commutation relations (\ref{3.3}) take the form
\begin{align}
\label{3.5}
b_r b^\dagger_s - b^\dagger_s b_r &= \delta_{rs} \, , \nonumber\\
b_r b_s - b_s b_r &= 0 \, , \nonumber\\
b^\dagger_r b^\dagger_s - b^\dagger_s b^\dagger_r & =  0 \, .
\end{align}
These equations are satisfied by (cf. also (\ref{2.11}) and (\ref{2.12}))
\be
\label{3.6}
b_r = e^{- \frac{i}{\hbar} \Theta_r} N^{1/2}_r \, , \quad \quad b^\dagger_r = N^{1/2}_r e^{\frac{i}{\hbar} \Theta_r} \, ,
\ee
where $N_r$  and $\Theta_r$ are Hermitian operators which satisfy the relation
\be
\label{3.7}
e^{- \frac{i}{\hbar} \Theta_r} f (N_r) = f (N_r + 1) e^{- \frac{i}{\hbar} \Theta_r} \, .
\ee
The eigenvalue of the operator $N_r$  are the integers $N'_r = 0, 1, 2, \ldots$.

$N_r = b^\dagger_r b_r$ is the number of particles in the $r-th$ state.
What is essentially new in the Dirac paper is that he considers the number $N_r$ of systems in the  $r-th$
 energy level as a canonical variable. In the space of such variables Dirac establishes the Schr\"odinger wave equation, i.e., 
the Schr\"odinger function becomes a function of the many 
$N_r$  and time $t$.
Let us see how it works. Consider an atom interacting with a radiation field. The Hamiltonian is given by
\be
\label{3.8}
H = H_0 + V \, ,
\ee
where $H$ is the Hamiltonian of the unperturbed atomic system which is under the influence of the perturbation $V$. 
The wave function of the total system satisfies the time-dependent Schr\"odinger equation						       
\be
i \hbar \frac{\partial}{\partial t} \langle \cdots, t | \rangle = \langle \cdots, t | H | \rangle= \langle \cdots, t | (  H_0 + V)| \rangle \, . \nonumber
\ee
We then project onto the basis-states of the unperturbed system $H_0$, which are labeled by $r$ with eigenvalues $W_r$
\begin{align}
\label{3.9}
i \hbar \frac{\partial}{\partial t} \langle \cdots, t | r \rangle_0 &= \langle \cdots; t | H_0 | r \rangle_0 = W_r \langle \cdots; t | r \rangle_0 \nonumber\\
\mbox{or} \quad i \hbar \frac{\partial \psi_r}{\partial t} &= H_0 \psi_r = W_r \psi_r \, .
\end{align}
Expanding the Schr\"odinger wave function $\langle \cdots; t | \rangle$ in terms of $\langle \cdots; t | r \rangle_0$
 we have
\begin{align}
\label{3.10}
\langle \cdots; t | \rangle&= \sli_r \langle \cdots; t | r \rangle_0 {}_0 \langle r | \rangle\nonumber\\
\mbox{or} \quad \psi (t) &= \sli_r a_r (t) \psi_r \, .
\end{align}
$a_r (t) = \langle \cdots; t | r \rangle_0$
is the transformation function.
$|a_r|^2$ is the probability of the system being in the state $r$  at time $t$. 
The probabilities are normalized so that $\sli_r | a_r |^2 = 1$. 
To apply the theory directly to an assembly of $N$ similar independent systems, we must multiply each of these $a_r$ by$ \sqrt{N}$ so as to make 
$\sli_r |a_r|^2 = N$.
As a result we obtain in $|a_r|^2$  the probable number of systems in the state $r$. Note that $|a_r|^2$ must be an integer.
From (\ref{3.9}) and (\ref{3.10}) we obtain the rate of change of the expansion coefficients $a_r$:
\be
\label{3.11}
i \hbar \frac{d a_r}{ dt} = \sli_s V_{rs} a_s
\ee
and that for the complex conjugate $a^*_r$
\be
\label{3.12}
- i \hbar \frac{d a^*_r}{d t} = \sli_s a^*_s V_{sr} \, .
\ee
If we regard $a_r$  and $i \hbar a^*_r$  as canonical variables, the equations (\ref{3.11}) and (\ref{3.12}) 
take the Hamiltonian form with the Hamiltonian function
\be
\label{3.13}
F_r = \sli_{r, s} a^*_r V_{rs} a_s
\ee
namely,
\be
\label{3.14}
\frac{d a_r}{dt} = \frac{1}{i \hbar} \frac{\partial F_1}{\partial a^*_r} \, , \quad \quad i \hbar \frac{d a^*_r}{dt} =
- \frac{\partial F_1}{\partial a_r} \, .
\ee
In order to express the complex canonical variables $a_r$  and $a^*_r$ in their real and imaginary parts, Dirac then introduced the real canonical variables
$N_r,  \phi_r$ by the transformation
\be
\label{3.15}
a_r = N^{1/2}_r e^{- \frac{i}{\hbar} \phi_r} \, , \quad \quad a^*_r = e^{\frac{i}{\hbar} \phi_r} N^{1/2}_r \, .
\ee
In terms of these variables the Hamiltonian  $F_1$  becomes
\be
\label{3.16}
F_1 = \sli_{r, x} V_{rs} N^{1/2}_r N^{1/2}_s e^{\frac{i}{\hbar} (\phi_r - \phi_s)} 
\ee
and the equations that determine the rate at which transitions occur have the canonical form
\be
\label{3.17}
\dot{N}_r = - \frac{\partial F_1}{\partial \phi_r} \, , \quad \quad 
\dot{\phi}_r =  \frac{\partial F_1}{\partial N_r} \, .
\ee
   Actually, Dirac found it more convenient to deal with the canonical variables							
\be
\label{3.18}
b_r  = a_r  e^{- \frac{i}{\hbar}  W_r t} \, , \quad \quad 
b^*_r  = a^*_r  e^{\frac{i}{\hbar}  W_r t} \, .
\ee
$W_r$  being the energy of the stationary state $r$. We have $|b_r|^2$ equal to $|a_r|^2$,
the probable number of systems in the state $r$. For $\dot{b}_r$ we find
\be
\label{3.19}
i \hbar \dot{b}_r = \sli_s H_{rs} b_s = W_r b_r + \sli_s v_{rs} b_s \, ,
\ee
where we put
\begin{align}
\label{3.20}
V_{rs} &= v_{rs} e^{\frac{i}{\hbar} (W_r - W_s)} \, , \nonumber\\
H_{rs} &= W_r \delta_{rs} + v_{rs} \, .
\end{align}
Taking the $b$'s canonical variables instead of the $a'$s means dealing with the Hamiltonian
\be
\label{3.21}
F = \sli_{r, s} b^*_r H_{rs} b_s \, ,
\ee						             												    
where we made a contact transformation	
\be
\label{3.22}
b_r = N^{1/2}_r e^{- \frac{i}{\hbar} \theta_r} \, , \quad \quad
b^*_r = N^{1/2}_r e^{ \frac{i}{\hbar} \theta_r}	
\ee										   						    
to the new canonical variables $N_r$   and  $\Theta_r$.
The Hamiltonian (\ref{3.21}) will now become
\be
\label{3.23}
F =  \sli_{r, x} H_{rs} N^{1/2}_r N^{1/2}_s e^{ \frac{i}{\hbar} (\Theta_r - \Theta_s)}  \, .
\ee
The equations for the rate of change of $N_r$    and  $\Theta_r$     will take the cononical form
\be
\label{3.24)}
\dot{N}_r = - \frac{\partial F}{\partial \Theta_r} \, \quad \quad
\dot{\Theta}_r = 
\frac{\partial F}{\partial N_r} \, .
\ee
The Hamiltonian may be written
\be
\label{3.25}
F = \sli_r W_r N_r + \sli_{r, s} v_{rs} N^{1/2}_r  N^{1/2}_s  e^{ \frac{i}{\hbar} (\Theta_r - \Theta_s)}  \, .
\ee
The first term, $\sli_r W_r N_r$, is the total proper energy of the assembly, and the second may be regarded as the additional energy due to perturbation.

Thus far Dirac had used only $c$-numbers. Now we assume the variables   $b_r, i \hbar b^*_r$
                  to be canonical $q$-numbers satisfying the quantum condition
\be
b_r i \hbar b^\dagger_r - i \hbar b^\dagger_r b_r = i \hbar \nonumber
\ee
or									
\be
\label{3.26}
b_r b^\dagger_r - b^\dagger_r b_r = 1
\ee             
and
\begin{align}
\label{3.27}
b_r b_s - b_s b_r & = 0 \, , \quad \quad b^\dagger_r b^\dagger_s - b^\dagger_s b^\dagger_r = 0 \, , \nonumber\\
b_r b^\dagger_s - b^\dagger_s b_r &= 0 \quad (s \neq r) \, .
\end{align}
The previous transformation equations must now be written in the form
\begin{align}
\label{3.28}
b_r & = e^{- \frac{i}{\hbar} \Theta_r} N^{1/2}_r
= (N_r + 1)^{1/2} e^{- \frac{i}{\hbar} \Theta_r}
	\, , \\
\label{3.29}
b^\dagger_r &= (N_r)^{1/2} e^{\frac{i}{\hbar} \Theta_r} = e^{\frac{i}{\hbar} \Theta_r}  (N_r + 1)^{1/2} 
\end{align}
in order that $N_r, \Theta_r$ be canonical variables satisfying the commutation rules
\be
[\Theta_r, N_r] = i \hbar \nonumber
\ee
or, equivalently, 	
\be
\label{3.30}
\left[ e^{\frac{i}{\hbar} \Theta_r}, N_r \right] = e^{\frac{i}{\hbar} \Theta_r} \, .
\ee
The equations (\ref{3.26}, \ref{3.27}, \ref{3.28}, \ref{3.29}) are precisely the same equations as the ones that are listed in (\ref{3.5}, \ref{3.6}, \ref{3.7}).
The Hamiltonian becomes
\begin{align}
\label{3.31}
F &= \sli_{r, s} b^\dagger_r H_{rs} b_s = \sli_{r, s} (N_r)^{1/2} e^{\frac{i}{\hbar} \Theta_r} H_{rs} (N_s + 1)^{1/2} e^{- \frac{i}{\hbar} \Theta_s} \nonumber\\
&= \sli_r W_r N_r + \sli_{r, s} v_{rs} (N_r)^{1/2}  (N_s + 1 - \delta_{rs})^{1/2} e^{\frac{i}{\hbar} (\Theta_r - \Theta_s)} 
\end{align}
and the Schr\"odinger equation can be written in terms of $N'_r$
   by noting that $e^{- \frac{i}{\hbar} \Theta_r}$  and  $e^{\frac{i}{\hbar} \Theta_r}$  
play the role of creation and annihilation operators. With the Schr\"odinger amplitude for observing particles at
$N'_1, N'_2, \ldots N'_r, \dots$
\be
\langle N'_1, N'_2, \ldots N'_r, \dots, t | \psi \rangle = \psi (N'_1, N'_2, \ldots N'_r, \ldots , t) \nonumber
\ee
we obtain		
\begin{align}
\label{3.32}
e^{- \frac{i}{\hbar} \Theta_r} \psi (N'_1, \ldots N'_r, \ldots, t) &= \psi (N'_1, \ldots, N'_r + 1, \ldots; t) \\
\label{3.33}
e^{\frac{i}{\hbar} \Theta_r} \psi (N'_1, \ldots N'_r, \ldots, t) &= \psi (N'_1, \ldots, N'_r - 1, \ldots; t) \, .
\end{align}
The evolution of $\psi (N'_1, \ldots, t)$ can thus be written as
\begin{align}
\label{3.34}
&  i \hbar \frac{\partial}{\partial t} \psi (N'_1, N'_2, \ldots; t) \nonumber\\
& = F \psi (N'_1, N'_2, \ldots; t) \nonumber\\
&= \sli W'_r N'_r \psi (N'_1, N'_2, \ldots; t) \nonumber\\
& + \sli_{r, s} v_{rs} (N'_r)^{1/2} 
(N'_s + 1 - \delta_{rs} )^{1/2} \psi  (N'_1, \ldots, N'_r - 1, \ldots, N'_s + 1, \ldots; t) \, .
\end{align}
Observe that so far we have a theory of free photons. This is because in the Lagrangian (\ref{3.21}) 
we see that each creation operator $b_s$ is accompanied by an annihilation operator $b^\dagger_r$; 
hence there is a continuous transition of photons from one mode to another. Hence there is no hope 
of calculating the probabilities for emission or absorption of photons by an atomic electron. Dirac 
cures this situation by appending an extra interaction term to the original Hamiltonian. It takes the form
\be
H_{in} = \sli_{r \neq 0} \lk v_r b^\dagger_r + v^*_r b_r \rk \, . \nonumber
\ee
With the use of such a linear interaction term, Dirac is able to compute Einstein's famous A and B coefficients after quantizing the
$(b^*_r, b_r)$ according to the $(N_r, \Theta_r)$ variables:
\be
H_{in} = \sli_r \lk v_r e^{\frac{i}{\hbar} \Theta_r} \sqrt{N_r + 1} + v^*_r e^{- \frac{i}{\hbar} \Theta_r} \sqrt{N_r} \rk \, . \nonumber
\ee
We will refrain from further elaboration of Dirac's second quantization prescription simply because his way of quantizing a field 
theory to become a theory of many particles which arise from excitations of underlying quantum fields is of mere historical 
interest now. The same fate was suffered by Dirac's hole theory, which was replaced by a highly successful theory of electrons 
and positrons, quantum electrodynamics. It is  here that the product of local second-quantized field operators, Jordan's invention, 
proved to be the correct instrument to handle with great success all the interaction processes where electrons, positrons and photons 
are involved.
\bi

\no
\section{Equivalence of Jordan's and Dirac's second quantization procedure}

Let's look at the relation between the usual one-particle Q.M. and the second quantization formalism.
 The connection is made at the level of the expectation value of a physical property $f$:
\be
\label{4.1}
\langle | f | \rangle = \sli_{b', b''} \langle | b' \rangle \langle b' | f | b'' \rangle \langle b'' | \rangle \, . 
\ee
$\langle b'' | \rangle$
     is the wave function of the state $| \rangle$     described with respect to the measurements $b''$. $\langle |b'\rangle$ 
 is the complex conjugate wave function described with respect to the measurements $b'$. 
From (\ref{4.1}) we obtain
\be
\label{4.2}
\langle | f | \rangle= \sli_{b', b''} \psi (b')^* \langle b' | f | b'' \rangle \psi (b''') \, .
\ee
This looks the same as the operator
\be
\label{4.3}
F = \sli_{b', b''} a (b')^\dagger \langle b' | f | b'' \rangle a (b'') \, .
\ee
Here $a(b')^\dagger$  and $a(b'')$ are operators. It is as though we had taken the numerical wave functions for a single system and extended them into operators, 
ending up with a description of a many-particle system. This is the origin of the name ``second quantization''.
To verify that our replacements
\begin{align}
\label{4.4}
\langle b'' | \rangle & \to a (b'') \, , \nonumber\\
\langle | b' \rangle & \to a (b')^\dagger
\end{align}
do not correspond to anything new but are merely a convenient description of $n$-particle 
systems (at the moment $n = 1$), let's consider an arbitrary linear operator $X$ acting on a one-particle state:
\be
\label{4.5}
X = \sli_{b', b''} | b' \rangle \langle b' | X | b'' \rangle \langle b'' | \, , \mbox{i.e.}, \quad X \, \mbox{in} \, b' \, \mbox{representation} \, .
\ee
Then
\be
X | \rangle_{old} = \sli_{b', b''} | b' \rangle \langle b' | X | b'' \rangle \langle b'' | \rangle_{old} \, , \quad \langle b'' | \rangle_{old} \, 
\mbox{wave function} \quad \psi (b'') \, . \nonumber
\ee
The one-particle state can be described in either way:
\be
\label{4.6}
| \rangle_{old} = \sli_{b'} | b' \rangle \langle b' | \rangle_{old} =  \sli_{b'} | b' \rangle \psi (b') 
\ee
or
\be
\label{4.7}
| \rangle_{new} = \sli_{b'''} a (b''')^\dagger | 0 \rangle \psi (b''') \, , \quad | b' \rangle= a (b')^\dagger | 0 \rangle \, .
\ee
Hence we obtain
\begin{align}
X | \rangle_{new} &= \sli_{b', b''} | b' \rangle \langle b' | X | b'' \rangle  \langle b'' | \sli_{b'''} a (b''')^\dagger | 0 \rangle \psi (b''') 
\nonumber\\
&= \sli_{b', b'', b'''} | b' \rangle \langle b' | X | b'' \rangle\langle 0 | a (b'') a (b''')^\dagger | 0 \rangle \psi (b''') \nonumber\\
& \hspace{2cm} a (b'') a     (b''')^\dagger = \delta_{b'' b'''} + a (b''')^\dagger a (b'') \nonumber\\
& \hspace{3cm} a (b'') | 0 \rangle = 0 \, , \langle 0 | 0 \rangle = 1 
\nonumber\\
&= \sli_{b', b''} | b' \rangle \langle b' | X | b'' \rangle \psi (b'') = \sli_{b', b''} | b' \rangle \langle b' | X | b'' \rangle \langle b'' | \rangle_{old}
  \nonumber\\
&= X | \rangle_{old} \, . \nonumber
\end{align}
Now let the first particle have value $b_1$, the secon $b_2$, etc. and let every mode $b_i$ contain only one particle. This $N$-particle basis state is then described by
\be
\label{4.8}
a (b_1)^\dagger a (b_2)^\dagger \ldots a (b_N)^\dagger  | 0 \rangle\, .
\ee
Generalizing (\ref{4.7}) to a general $N$-particle state $| \phi' \rangle$         in which the particles have a wave function  $\varphi (b_1, b_2, \ldots, b_N)$
we simply write a linear combination of states (\ref{4.8}):
\be
\label{4.9}
| \phi' \rangle = \sli_{b_1, \ldots, b_N} \varphi (b_1, b_2, \ldots, b_N) a^\dagger(b_1) a^\dagger (b_2) \ldots a^\dagger (b_N) | 0 \rangle \, .
\ee
If $\varphi$    is properly symmetrized, then the amplitude for observing one particle in mode $b'_1$, one in mode $b'_2$, etc. is given by
\be
\langle b'_1, b'_2, \ldots, b'_N | \phi' \rangle \quad \mbox{not yet symmetrized}. \nonumber
\ee
We want to check that the second quantized version of Jordan's gives the same result as the one obtained by our old Schr\"odinger formulation.               
To show this, take another state
\be
\label{4.10}
| \tilde{\phi}' \rangle= \sli_{b'_1 \ldots, b'_N} \tilde{\varphi} (b'_1, b'_2, \ldots, b'_N) a^\dagger (b'_1) \ldots a^\dagger (b'_N) | 0 \rangle
\ee
and calculate the inner product
\begin{align}
\label{4.11}
\langle \tilde{\phi'} | \phi' \rangle & = \sli_{b_1, \ldots, b_N \atop b'_1, \ldots b'_N} \tilde{\varphi}^* (b'_1, \ldots, b'_N) \varphi (b_1, \ldots, b_N) 
\nonumber\\
& \langle 0 | a (b'_N) \ldots
a (b'_1) a^\dagger (b_1) \ldots a^\dagger (b_N) | 0 \rangle \, .
\end{align}
Now we are going to apply the quantization relations (C.R.'s) $[a (b'), a (b)^\dagger] = \delta_{b' b}$ 
 to bring the $a'$s to the right where they can act on  $| 0 \rangle, a |   0 \rangle = 0$.

To evaluate the vauum expectation value in (\ref{4.11}), let's forget the problem for the time being and consider the function
\be
\label{4.12}
e^{\sli_b \alpha (b) a^\dagger (b)}, \alpha (b) \quad \mbox{parametric    function} \, .
\ee
Now take this expression and differentiate:
\be
\frac{\partial}{\partial \alpha (b_1)} \frac{\partial}{\partial \alpha (b_2)} \ldots  \frac{\partial}{\partial \alpha (b_N)}
\lk e^{\sli_b \alpha (b) a^\dagger (b)} \rk_{\alpha = 0} \equiv a^\dagger (b_1) a^\dagger (b_2) \ldots a^\dagger (b_N) \, . \nonumber
\ee 
In the same way
\be
\frac{\partial}{\partial \alpha' (b'_1)} \frac{\partial}{\partial \alpha' (b'_2)} \ldots  \frac{\partial}{\partial \alpha' (b'_N)}
\lk e^{\sli_{b'} \alpha' (b') a (b')} \rk_{\alpha' = 0} \equiv a (b'_1) a (b'_2) \ldots a (b'_N) \, . \nonumber
\ee
Then
\begin{align}
\langle \tilde{\phi}' | \phi' \rangle & = \sli_{b_1, \ldots b_N \atop b'_1, \ldots b'_N} \tilde{\phi}^* \lk b'_1, \ldots, b'_N \rk \varphi (b_1,  \ldots, 
b_N) 
\frac{\partial}{\partial \alpha' (b'_1)} \ldots 
\frac{\partial}{\partial \alpha' (b'_N)} 
\frac{\partial}{\partial \alpha (b_1)} \ldots
 \frac{\partial}{\partial \alpha (b_N)}\nonumber\\
& \cdot
\langle 0 | e^{\sli_{b'} \alpha' (b') a (b')} e^{\sli_b \alpha (b) a^\dagger (b)} | 0 \rangle_{\alpha = \alpha' = 0} \nonumber
\end{align}
To compute the quantity $\langle 0 | \cdots | 0 \rangle$                      we write
\begin{align}
\label{4.13}
& \langle 0 | e^{\sli_{b'} \alpha' (b') a (b') \sli_b \alpha (b) a^\dagger (b)} e^{- \sli_{b'} \alpha' (b') a (b')}  e^{\sli_{b'} \alpha' (b') a (b')} | 0 \rangle
\nonumber\\
& \hspace{2cm}  
S F (a^\dagger) S^{- 1} = F (S a^\dagger S^{- 1}) \nonumber\\ 
& \hspace{2cm} \mbox{just write in power series:} \hspace{1cm} S (a^\dagger)^N S^{- 1} 
= S a^\dagger S^{- 1} S \ldots S a^\dagger S^{- 1} \nonumber \\
& \hspace{4cm} = (S a^\dagger S^{- 1})^N \nonumber\\
&= \langle 0 | e^{\sli_{b'} \alpha (b') \alpha (b')} \sli_b \alpha (b) a^\dagger (b) e^{- \sli_{b'} \alpha' (b') a (b')} | 0 \rangle \, .
\end{align}
To proceed, we have to know
\be
\label{4.14}
e^{\sli_{b'} \alpha' (b') a (b')} a^\dagger(b) 
e^{\sli_{b'} \alpha' (b') a (b')} 
\equiv F \{ \alpha' \} \, .
\ee
Taking the derivative of this expression, we first set up a differential equation:
\begin{align}
& \frac{\partial}{\partial \alpha' (b'')} \lk 
e^{\sli_{b'} \alpha' (b') a (b')} a^\dagger (b)
e^{- \sli_{b'} \alpha' (b') a (b')} \rk \nonumber\\
&= e^{ \sli_{b'} \alpha' (b') a (b')} \underbrace{\lk a (b'') a^\dagger (b) - a^\dagger (b) a (b'') \rk}_{= \delta_{b b''} }
e^{- \sli_{b'} \alpha' (b') a (b')} \nonumber\\
&= \delta_{b b''} = \frac{\partial}{\partial \alpha' (b'')} (\alpha' (b)) \, . \nonumber
\end{align}
From the last equation we obtain
\be
F (\alpha') = e^{\sli_{b'} \alpha' (b') a (b')} 
a^\dagger (b) 
e^{- \sli_{b'} \alpha' (b') a (b')} = \alpha' (b) + const. \nonumber
\ee
and since $F \{ \alpha' (b') \} - \alpha' (b)$ is independent of $\alpha'$, 
we can determine its value at $\alpha'= 0$ to leave us with the constant of integration $a^\dagger (b)$.
From (\ref{4.14})  follows
\be
\label{4.15}
e^{\sli_{b'} \alpha' (b') a (b')} 
a^\dagger (b) 
e^{- \sli_{b'} \alpha' (b') a (b')} = \alpha' (b) + a^\dagger(b)
\ee
and from (\ref{4.13}):
\be
\langle 0 | e^{\sli_b \alpha (b)  \lk a^\dagger (b) + \alpha' (b) \rk} | 0 \rangle = e^{\sli_b \alpha (b) \alpha' (b)} \, .\nonumber
\ee
Thus, the scalar product looks like
\be
\label{4.16}
\langle \tilde{\phi'} | \phi' \rangle =
\left. \sli_{b_1, \ldots, b_N \atop
b'_1, \ldots, b'_N} \tilde{\varphi}^* (b'_1, \ldots, b'_N) \varphi (b_1, \ldots, b_N) \frac{\partial}{\partial \alpha' (b'_1)} \ldots 
\frac{\partial}{\partial \alpha (b_N)} 
e^{\sli_b \alpha (b) \alpha' (b)} \right|_{\alpha = \alpha' = 0} \, .
\ee
When the first $N$ derivatives with respect to  $\alpha (b_i)$ are performed, we are left with
\begin{align}
& \langle \tilde{\phi}' | \phi' \rangle = 
 \sli_{b_1, \ldots, b_N \atop
b'_1, \ldots, b'_N} \varphi^* (b'_1, \ldots, b'_N) \varphi (b_1, \ldots, b_N) \nonumber\\
& \frac{\partial}{\partial \alpha' (b'_1)} \ldots
\frac{\partial}{\partial \alpha' (b'_N)} \left. 
\lk \alpha' (b_1) \ldots \alpha' (b_N) \rk \right|_{\alpha' = 0} \nonumber
\end{align}
\be
\varphi \, \mbox{symmetric function} \, = N! \sli_{b_1, b_2, \ldots b_N} \varphi^* (b_1, \ldots, b_N) \varphi (b_1, \ldots, b_N) \, , \nonumber
\ee
which, after redefinition of      $| \phi \rangle$:
\be
\label{4.17}
| \phi \rangle = \frac{1}{\sqrt{N!}} \sli_{b_1, \ldots, b_N} \varphi (b_1, \ldots, b_N) a^\dagger (b_1) \ldots a^\dagger (b_N) | 0 \rangle
\ee
yields just the old version of the scalar product. 
Now the amplitude for observing particles at $b'_1, b'_2, \ldots, b'_N$ is given by
\begin{align}
\label{4.18}
\langle b'_1, \ldots, b'_N | \phi \rangle &= \frac{1}{\sqrt{N!}} \langle 0 | a (b'_1) \ldots a (b'_N) | \phi \rangle = \varphi (b'_1, \ldots, b'_N) \\
\label{4.19}
\langle \phi | \phi \rangle &= 1 \\
\label{4.20}
| \phi \rangle &= \sli_{b_1, \ldots, b_N} | b_1, \ldots, b_N \rangle\langle b_1, \ldots, b_N | \phi \rangle\, .
\end{align}
In other words, the operator
\be
\label{4.21}
1 = \sli_{b_1, \ldots, b_N} | b_1, \ldots, b_N \rangle \langle b_1, \ldots, b_N | 
\ee
is the unit operator when operating on properly symmetrized $N$-particle states.

Finally we come to time-dependent problems.
Given H, then the equation of motion for the operator $X(t)$ is given by  $(\hbar = 1)$
\be
\label{}
\frac{d}{dt} X (t) = \frac{1}{i}  [X (t), H] \nonumber
\ee
with the solution
\be
X (t) = e^{i H t} X e^{-  i H t} \, . \nonumber
\ee
If the single particle Hamiltonian is given by $H = \frac{\vp^2}{2m}$, then in a representation where $\vp$  is diagonal, we obtain
\begin{align}
H &= \sli_ {\vp} a^\dagger (\vp) \frac{\vp^2}{2m} a (\vp) \nonumber\\
\frac{d}{dt} a (\vp, t) &= \frac{1}{i} [a (\vp, t),  H (t)] = - i \frac{\vp^2}{2m} a (\vp, t) \nonumber\\
a (\vp, t) &= e^{-i \frac{\vp^2}{2m} t} a (\vp) \, , \quad \quad a (\vp) \equiv a (\vp, t = 0) \, . \nonumber
\end{align}
Likewise
\be
a^\dagger (\vp, t) = e^{i \frac{\vp^2}{2m} t} a^\dagger (p) \, . \nonumber
\ee
In general, for a 2nd quantized operator $X$,
\be
X = \sli_{\vp, \vp'} a^\dagger (\vp) \langle \vp | X | \vp' \rangle a (\vp') \, , \nonumber
\ee
we have the time dependence
\be
X (t) = \sli_{\vp, \vp'} a^\dagger (\vp, t) \langle \vp | X | \vp' \rangle a (\vp', t) \, , \nonumber
\ee
which satisfies
\be
\frac{d X}{d t} = \frac{1}{i} [X, H] \, .
\nonumber
\ee
For the operator $H = \frac{\vp^2}{2m}$ we have the equivalent expressions:
\begin{align}
X (t) &= \sli_{\vp, \vp'} a^\dagger (\vp) e^{i \frac{\vp^2}{2m} t} \langle \vp | X | \vp' \rangle 
e^{- i \frac{\vp'^2}{2m} t} a (\vp') \nonumber\\
&= \sli_{\vp, \vp'} a^\dagger (\vp)  \langle p | e^{i H t} X e^{-  H t} | \vp' \rangle a (\vp')\nonumber\\
&= \sli_{\vp, \vp'} a^\dagger (\vp)  \langle \vp | X (t) | \vp' \rangle a (\vp') \, . \nonumber
\end{align}
Now we want to switch to the coordinate basis. Here we write instead of $a(\vp)$ the operator $ a(\vx) \equiv \psi (\vx)$.

Since
\be
| \vx \rangle = \sli_{\vp} | \vp \rangle \langle \vp | \vx \rangle= \sli_{\vp} | \vp \rangle \frac{e^{- i \vx \cdot \vp}}{\sqrt{V}}
\nonumber
\ee
we obtain
\be
\psi^\dagger (\vx) = \sli_{\vp} a^{\dagger} (\vp) \frac{1}{\sqrt{V}} e^{- i \vx \cdot \vp} \nonumber
\ee
and
\be
\psi (\vx) = \sli_{\vp} \frac{1}{\sqrt{V}} e^{i \vx \cdot \vp} a (\vp) \, .\nonumber
\ee
Therefore
\begin{align}
[\psi (\vx), \psi^\dagger (\vy)] &= \sli_{\vp, \vp'} \frac{1}{V} [a (\vp), a^\dagger (\vp')] e^{i \vx \cdot \vp} 
e^{- i \vy \cdot \vp'} \nonumber\\
&= \sli_{\vp} \frac{1}{V} e^{i \vp \cdot (\vx - \vy)} \raisebox{-1ex}{$\overrightarrow{V \to \infty}$} \il \frac{d^3 \vp}{(2 \pi)^3}
e^{i \vp \cdot (\vx - \vy)} = \delta^3 (\vx - \vy) \, . \nonumber
\end{align}
Now the operator equation of motion is given by
\be
i \frac{\partial}{\partial t} \psi (\vx, t) = [\psi (\vx, t), H ] \nonumber
\ee
Substitute
\begin{align}
\psi (\vx, t) &= \sli_{\vp} \frac{1}{\sqrt{V}} a (\vp, t) e^{i \vp \cdot \vx} \nonumber\\
&= \sli_{\vp} \frac{1}{\sqrt{V}} a (\vp) e^{i (\vp \cdot \vx - \frac{\vp^2}{2m} t)} \nonumber\\
&= \sli_{\vp} \langle \vx, t | \vp \rangle a (\vp) \nonumber
\end{align}
to obtain
\be
i \frac{\partial}{\partial t} \psi (\vx, t) = - \frac{1}{2m} \vec{\nabla}^2 \psi (\vx, t) \nonumber
\ee
(Recall the Schr\"odinger equation for the amplitude $\langle \vx, t | \vp \rangle$:
\begin{align}
i \frac{\partial}{\partial  t} \langle \vx, t | \vp \rangle &= \langle \vx, t | H | \vp \rangle = \langle \vx, t | \frac{\vp^2}{2m} | \vp \rangle \nonumber\\
&= \sli_{\vx'} \langle \vx, t | \frac{\vp^2}{2m} |  \vx', t \rangle \langle \vx', t| \vp \rangle = - \frac{1}{2m} \vec{\nabla}^2 
\langle \vx, t | \vp \rangle \, . \nonumber
\end{align}
But in the present chapter, $\psi$  denotes an operator.  $\psi (\vx, t)$ destroys a particle at place $\vx$ at time $t$.

\section*{Acknowledgement}

I enjoyed innumerable discussions with Nils Schopohl on the early days of quantum mechanics.

\end{document}